\documentstyle[aps,prb,twocolumn]{revtex}

\begin{document}

\preprint{PRB (in press)}
\draft

\wideabs{
\title{Range dependence of interlayer exchange coupling}

\author{Peter M. Levy}

\address{Department of Physics, New York University, 4 Washington Place, 
New York, NY 10003 USA}

\author{Sadamichi Maekawa}

\address{Institute for Materials Research, Tohoku University, Sendai 
980-77, Japan}

\author{Patrick Bruno\cite{new-address}}

\address{Institut d'\'Electronique Fondamentale, Universit\'e Paris-Sud,
B\^atiment 220, F-91405 Orsay, France}

\date{30 January 1998; revised 8 April 1998}

\maketitle

\begin{abstract}
We have considered the effects of non magnetic impurities and interface
roughness on the interlayer coupling between magnetic layers in metallic
multilayers. The two types of defects alter the interlayer coupling in
quite different ways. Elastic electron scattering by impurities in the
non magnetic spacer layers between magnetic layers produces an
exponential decay of the coupling with a characteristic decay length
that is considerably longer than the ``global'' transport mean free path
for the spacer layer with its surrounding interfaces. Interfacial
roughness leads to an attenuation of the coupling that is related to the
width of the roughness in relation to the Fermi wavelength;
roughness does not alter the range dependence of the coupling. For
certain types of electrical transport, e.g., for current perpendicular to
the plane of the layers, the scattering from interface roughness and
impurities in the spacer layers contribute on an equal footing to the
exponential decay of the electron propagators, i.e., global mean free
path. We show that interface roughness and impurities in the spacer layer 
affect the interlayer coupling differently.

\end{abstract}


\pacs{Published in: Phys. Rev. B {\bf 58}, 5588--5593 (1998)}
} 

\section{Introduction}\label{sec:intro}

The range dependence of the indirect coupling between magnetic ions mediated by 
conduction electrons, i.e., the Ruderman-Kittel-Kasuya-Yosida (RKKY) 
interaction, in the presence of scattering by non magnetic impurities has been 
debated over the past 35 years.\cite{ deGennes1962, deChatel1981, 
Bulaevskii1986, Zyuzin1986, Jagannathan1988} Recently this question has been 
resolved; it is now clear that scattering by non magnetic impurities does not 
alter the range dependence of the coupling between two magnetic ions; it only 
introduces a phase  factor that depends on the specific distribution of 
impurities between the two ions.\cite{ Bulaevskii1986, Zyuzin1986, 
Jagannathan1988} As long as  one does {\em not\/} average over different 
impurity configurations the coupling between a pair of magnetic ions remains 
undamped. Only when one looks at properties that require an average of the 
coupling between two ions for different realizations of impurity distributions, 
and therefore by taking an average over the phase factor  produced by the 
impurity scattering, does one find that the coupling is damped.\cite{ 
deGennes1962}

The recent interest in the interlayer exchange coupling in magnetic 
multilayers\cite{ Parkin1990} raises the questions whether scattering in the 
non magnetic spacer layers dampens the coupling between magnetic layers, and what 
effect the roughness of  interfaces will have on the range of the coupling. Here 
we show that for the coupling between two {\em sheets\/} of spins one does 
average over different realizations of impurity distributions in the intervening 
spacer, and the coupling is exponentially damped as a function of the distance 
between planes, i.e., magnetic layers. The damping is proportional to the 
strength of the impurity scattering in the spacer; it has {\em nothing\/} to do 
with the scattering due to roughness at the interfaces. While this 
characteristic decay distance of the interlayer coupling and  the transport mean 
free path in the spacer layer\cite{ Valet1993} are both due to impurity 
scattering, they are in {\em no\/} way simply related to one another. 
Interfacial roughness attenuates the coupling\cite{ Wang1990} in proportion to 
the size of the inter diffused region relative to the Fermi wavelength of the 
conduction electrons providing the coupling. However, this decrease does {\em 
not\/} alter the range dependence of the coupling. For this reason it would be 
completely erroneous to combine the effects of the scattering from impurities 
in the spacer and interface roughness into one decay coefficient to produce an 
exponential decay of the interlayer coupling. While this 
procedure is correct for certain types of  electrical transport in magnetic 
multilayers, e.g.,  for current perpendicular to the plane of the layers 
(CPP),\cite{ Zhang1991} interlayer coupling is not in this class of situations. 
Therefore the range of the interlayer coupling is considerably {\em longer\/}  
(characteristic decay of the coupling is considerably slower) than what one 
anticipates from the transport properties (resistivity) of a magnetic 
multilayer.

In the following section we first derive the RKKY coupling between a pair of 
magnetic ions in the presence of non magnetic impurities. Next we show how 
averaging over the phase induced by the impurity scattering produces a coupling 
that decays with the distance between the ions, and the extent to which this 
can be described by an exponential of the distance. We use the approximation of 
representing the coupling between two magnetic layers as that due to the two 
planes of magnetic ions that interface with the nonmagnetic spacer layer,\cite{ 
Yafet1987} and we show that when one averages over the magnetic ions in these 
planes that this coupling has a slowly decaying exponential component as a 
function of the  thickness of the spacer layer. Finally we take into account of 
effect of the roughness of the interfaces  and show how this attenuates the 
coupling.

\section{Two Magnetic Ions}\label{sec:ions}

To calculate the coupling between two magnetic ions when the conduction 
electrons are scattered by impurities we use the approach adopted by Bulaevskii 
and Panyukov.\cite{ Bulaevskii1986} It consists of using a semiclassical form 
for the electron propagator (Green's function) connecting the positions of the 
two ions, and taking  account of the scattering by impurities through the phase 
of the propagator. The coupling between two magnetic impurities located, 
respectively, at ${\bf r}$ and ${\bf r}^\prime$ is given by
\begin{equation}\label{eq:J-GG}
J({\bf r}, {\bf r}^\prime ) \sim \int_{\varepsilon_F - {\rm i}\infty}^{
\varepsilon_F + {\rm i}\infty} \! dz \ 
{\rm Tr} \left[ G ({\bf r}, {\bf r}^\prime ; z ) 
G ({\bf r}^\prime , {\bf r}; z ) \right] .
\end{equation}
Here, $G$ is the Green's function corresponding to a particular configuration 
of impurities. A typical Feymann diagram contributing to the above expression, 
is a ``bubble'' diagramm with an electron line going from ${\bf r}$ to ${\bf 
r}^\prime$ over a given set of impurities, and then back from ${\bf r}^\prime$ 
to ${\bf r}$ oven a (generally) different set of impurities; this contribution  
to Eq.~(\ref{eq:J-GG}) contains a phase factor
\begin{equation}
\exp \left\{ {\rm i}k(z)  \left[ L({\bf r} \to {\bf r}^\prime ) + L({\bf 
r}^\prime \to {\bf r} ) \right] \right\} ,
\end{equation}
where $k(z)$ is the (complex) wavevector corresponding to the complex energy 
$z$, and where $L({\bf r} \to {\bf r}^\prime )$ ($L({\bf r}^\prime \to {\bf r} 
)$) is the length of the path from ${\bf r}$ to ${\bf r}^\prime$ (${\bf 
r}^\prime$ to ${\bf r}$) over the corresponding set of impurities. 

For points ${\bf r}$ and ${\bf r}^\prime$ not very close to each other, when 
summing over all diagrams (i.e., over all paths over impurities), the above 
factor oscillates rapidly, leading to a strong cancellation. Thus, as pointed 
out by Bulaevskii and Panyukov,\cite{ Bulaevskii1986} the only significant terms 
are due to the paths going through impurities lying on a straight line between 
${\bf r}$ and ${\bf r}^\prime$ (the impurities being passed sequentially, 
without back tracking); for all such paths, the phase factor is
\begin{equation}
\exp \left[ 2{\rm i}k(z) |{\bf r} - {\bf r}^\prime | \right] .
\end{equation}

We wish to stress that the present situation is completely different from the 
one encountered when computing the two-point conductivity $\sigma ({\bf r}, 
{\bf r}^\prime )$, which involves the product $G ({\bf r}, {\bf r}^\prime ; 
\varepsilon_F + {\rm i}0^+ ) 
G ({\bf r}^\prime , {\bf r}; \varepsilon_F - {\rm i}0^+ )$. This yields an 
oscillatory factor
\begin{equation}
\exp \left\{ {\rm i}k_F  \left[ L({\bf r} \to {\bf r}^\prime ) - L({\bf 
r}^\prime \to {\bf r} ) \right] \right\} .
\end{equation}
Because of the minus sign in the above expression, there is a significant 
contribution from all diagrams such that the path ${\bf r}^\prime \to {\bf r}$ 
is the reverse of the path ${\bf r} \to {\bf r}^\prime$, so that $L({\bf r} \to 
{\bf r}^\prime ) = L({\bf r}^\prime \to {\bf r} )$; these are the ladder 
diagrms characteristic of a diffusive process, which give the leading 
contribution to $\sigma ({\bf r},{\bf r}^\prime )$.

Following Bulaevskii and Panyukov,\cite{ Bulaevskii1986} we obtain the 
expression of the coupling:
\begin{equation}\label{eq:J}
J({\bf r}, {\bf r}^\prime ) \sim \frac{\cos [2k_F | {\bf r} - {\bf r}^\prime | 
+\phi ({\bf r}, {\bf r}^\prime ) ]}{|{\bf r} - {\bf r}^\prime |^{3}} ;
\end{equation}
the phase shift due to impurity scattering is
\begin{equation}\label{eq:phi}
\phi ({\bf r}, {\bf r}^\prime ) = \frac{-2}{\hbar v_F} \int_0^{|{\bf r} - {\bf 
r}^\prime |} \! ds \, U({\bf r} + {\bf \hat n} s) ,
\end{equation}
where ${\bf \hat n}$ is a unit vector of the $({\bf r}, {\bf r}^\prime )$ axis 
and $U({\bf r})$ 
is the impurity scattering potential. Equations (\ref{eq:J}) and (\ref{eq:phi}) 
are valid if the perturbation potential $U({\bf r})$ is small compared to the 
Fermi energy. As there is a specific distribution of impurities between a pair 
of ions there is a definite phase; the coupling is phase shifted but it is 
{\em not\/} damped by impurity scattering (its decay law is as $|{\bf r} - {\bf 
r}^\prime|^{-3}$ like in the pure system). 

The distribution of $J({\bf r}, {\bf r}^\prime )$ is thus determined by the 
distribution of the phases $\phi ({\bf r}, {\bf r}^\prime )$. To ascertain the 
distribution of phase angles we fill the space between two spins with cubes 
of length $a$, and rewrite the phase integral Eq.~(\ref{eq:phi}) as a sum over 
the $N$ cells ($N\equiv |{\bf r} -{\bf r}^\prime | / a$) crossed by the 
trajectory
\begin{eqnarray}\label{eq:phi-i}
\phi({\bf r}, {\bf r}^\prime ) &=& \sum_{i=1}^N \phi_i \ ; \\
\phi_i &=& \left\{
\begin{array}{cl}
-2Ua/\hbar v_F \ \ & \mbox{(if impurity in cell $i$)} \\
0                 & \mbox{(if no impurity in cell $i$)}
\end{array} \right. \nonumber
\end{eqnarray}
If an impurity is in a cell it yields a contribution $\kappa = -2Ua/\hbar v_F = 
-(3 \pi)^{1/3} U/E_F$ to the phase; otherwise it gives zero. For a given 
concentration of impurities $c$ there is a probability $c$ at each site of 
their being an impurity and therefore of picking up a phase of $\kappa$, and a 
probability of $1-c$ of picking up zero. Therefore the phase in 
Eq.~(\ref{eq:phi-i}) is just a binomial distribution of $N$ events. 

>From the distribution of phases, one can compute the various moments of the 
exchange interaction distribution,
\begin{equation}
\overline{J^n}(R) \equiv \left< J^n({\bf r}, {\bf r}^\prime ) \right> ,
\end{equation}
where the angular brackets indicate that we average over all possible 
configurations of impurities. The averaging restores the translational 
and rotational invariance, therefore the moments $\overline{J^n}$ depend only 
on the distance $R \equiv |{\bf r} - {\bf r}^\prime |$ between the two spins. 
As was emphasized by various authors,\cite{ deChatel1981, Bulaevskii1986, 
Zyuzin1986, Jagannathan1988} the first moment $\overline{J}(R)$ bears little 
physical significance in the case of magnetic ions embedded in a disordered 
nonmagnetic host; for instance, the transition temperature of spin glasses is 
determined by the second moment $\overline{J^2}(R)$, not by the first moment. 
To illustrate this point, we compute now the first two moments, $\overline{J}(R)$ 
and $\overline{J^2}(R)$.

The configuration averaged exchange interaction (first moment) is given by
\begin{equation}\label{eq:J-av1}
\overline{J}(R) \sim \frac{{\rm Re} \left( {\rm e}^{2{\rm i}k_F R}  \left< {\rm 
e}^{{\rm i}\phi({\bf r}, {\bf r}^\prime )}\right> \right)}{R^3}.
\end{equation}
The characteristic function  of the phase,\cite{ vanKampen1981} for the binomial 
distribution is given as\cite{ Abramovitz1964} 
\begin{equation}
\left< {\rm e}^{{\rm i}\phi({\bf r}, {\bf r}^\prime )} \right> = \left[ (1-c) + 
c  {\rm e}^{{\rm i}\kappa} \right]^{R/a} ,
\end{equation}
where we have used $R=Na$. 

By taking the logarithm of the characteristic function for the binomial 
distribution we find in the limit of {\em low\/}  impurity concentrations 
\begin{eqnarray}\label{eq:exp-av}
\left< {\rm e}^{{\rm i}\phi({\bf r}, {\bf r}^\prime )}\right> &\approx& \exp 
\left[ c \left( {\rm e}^{{\rm i}\kappa} -1 \right) R/a \right] \nonumber \\
&=& {\rm e}^{{\rm i}c (\sin \kappa ) R/a} \, {\rm e}^{-c(1-\cos \kappa ) R/a} .
\end{eqnarray}
Thus, we obtain an exponential decay of the exchange interaction with a decay 
length $\lambda$ given by
\begin{equation}\label{eq:decay}
\lambda^{-1} = \frac{c}{a} \left( 1 -\cos\kappa \right) \approx  \frac{c\, 
\left( 3\pi^2\right)^{2/3}}{2a} \left( \frac{U}{\varepsilon_F} \right)^2
\end{equation}
and a shift in the Fermi wavevector
\begin{equation}\label{eq:shift}
\delta k = \frac{ c \ \sin \kappa }{2a} \approx  \frac{-c \, \left( 
3\pi^2\right)^{1/3} }{2a}  \left(\frac{U}{\varepsilon_F} \right) .
\end{equation}
The wavevector shift $\delta k$ is simply due to the shift in the average value 
of the potential; in the following it will be incorporated into a redefinition 
of the Fermi wavevector $k_F$.

In this way we find that the average RKKY coupling is
\begin{equation}\label{eq:J-av}
\overline{J}(R) \sim \frac{ \cos \left( 2 k_F R \right)\ {\rm e}^{-R/ 
\lambda}}{R^3} .
\end{equation}
This result was first obtained by de~Gennes.\cite{ deGennes1962} 

The second moment is easily calculated directly from Eq.~(\ref{eq:J}) and we 
obtain\cite{ Bulaevskii1986, Zyuzin1986, Jagannathan1988}
\begin{equation}
\overline{J^2}(R) \sim \frac{1}{2}\ \frac{1}{R^6} .
\end{equation}
Thus it appears clearly that
\begin{equation}
\frac{\overline{J^2}(R) - \overline{J}^2(R)}{\overline{J}^2(R)} \sim {\rm 
e}^{2R/\lambda } ,
\end{equation}
i.e., that the exchange interaction between a pair of magnetic ions is {\em 
not\/} a self-averaging quantity (in the sense of Kohn and Luttinger\cite{ 
Kohn1957}).

In the case of a system with a non-spherical Fermi surface, the configuration 
averaged exchange interaction takes a form similar to Eq.~(\ref{eq:J-av}), but 
the wavevector of oscillations and the decay length both depend on the direction 
${\bf \hat n}$. We stress that the decay length of the exchange interaction 
corresponding to a particular direction generally has no simple relation with 
the transport mean free path, because the later results from averaging over all 
directions.

\section{Two Sheets of Spins}\label{sec:sheets}

\subsection{Perfectly flat layers}

We now consider the interlayer exchange coupling between two ferromagnetics 
layers, $F_1$ and $F_2$. These are modelled by taking two sheets of magnetic 
ions of normal coordinates $r_{\bot 1}$ and $r_{\bot 2}$, respectively. Within 
a given sheet, we assume that all the magnetic moments are maintained parallel to 
each other by some intralayer exchange coupling (which we do not describe 
explicitely here); thus, the only variable is the angle between the 
magnetizations of the two sheets.

Following Yafet,\cite{ Yafet1987} we express the coupling between $F_1$ and 
$F_2$, as the sum over the pairs of magnetics ions $({\bf r}_1 ,{\bf r}_2 )$ 
(divided by the total area $S$), with ${\bf r}_1 \equiv ({\bf r}_{\| 1}, 
r_{\bot 1})$ belonging to $F_1$ and ${\bf r}_2 \equiv ({\bf r}_{\| 2}, 
r_{\bot 2})$ belonging to $F_2$:
\begin{equation}
I(r_{\bot 1}, r_{\bot 2}) \equiv \frac{1}{S} \int\! d^2{\bf r}_{\| 1} \int\! 
d^2{\bf r}_{\| 2} \ J({\bf r}_1 , {\bf r}_2 ) .
\end{equation}
To compute this, we first sum over all pairs $({\bf r}_1,{\bf r}_2)$ 
with ${\bf r}_1 - {\bf r}_2$ parallel 
to a given direction, i.e., we rewrite the above equation as 
\begin{equation}
I(r_{\bot 1}, r_{\bot 2}) = \int\! d^2\mbox{\boldmath $\rho$}_\| \ K(r_{\bot 1}, 
r_{\bot 2}; \mbox{\boldmath $\rho$}_\| ) ,
\end{equation}
with
\begin{equation}
K(r_{\bot 1}, r_{\bot 2}; \mbox{\boldmath $\rho$}_\| ) \equiv \frac{1}{S} \int\! 
d^2{\bf r}_\| \ J\left( ({\bf r}_\| , r_{\bot 1}), ({\bf r}_\| + \mbox{\boldmath 
$\rho$}_\| , r_{\bot 2}) \right) .
\end{equation}
It is easy to see that, when summing over ${\bf r}_\|$, all configurations of 
impurities between ${\bf r}_1 = ({\bf r}_\| , r_{\bot 1})$ and ${\bf r}_2 = 
({\bf r}_\| + \mbox{\boldmath $\rho$}_\|, r_{\bot 2})$ are encountered, thus 
this is equivalent to performing a configuration average, i.e.,
\begin{equation}
K(r_{\bot 1}, r_{\bot 2}; \mbox{\boldmath $\rho$}_\| ) = \overline{J}(R_{1,2}) ,
\end{equation}
where
\begin{equation}
R_{1,2} \equiv \sqrt{\mbox{\boldmath $\rho$}_\|^2 + D^2}
\end{equation} 
is the distance between the spins for the pairs considered, and $D \equiv 
|r_{\bot 1} - r_{\bot 2}|$ is the distance between the two sheets of spins. 

This implies that, in contrast to the exchange interaction between two magnetic 
ions, the exchange coupling between two sheets of spins is self-averaging in the 
sense of Kohn and Luttinger.\cite{ Kohn1957} Thus, it is only a function of the 
distance $D$ between the two sheets, and not of $r_{\bot 1}$ and $r_{\bot 2}$ 
separately.

The remaining integration over $\mbox{\boldmath $\rho$}_\|$ is then easily 
calculated by using Yafet's method,\cite{ Yafet1987} and we find
\begin{equation}
I(D) \sim 2\pi\, \mbox{Re} \left[ \frac{-{\rm i}}{\gamma}\ \frac{{\rm e}^{{\rm 
i}\gamma D}}{D^2} \right] ,
\end{equation}
with $\gamma = 2k_F + {\rm i}\lambda^{-1}$; finally, we obtain
\begin{equation}
I(D) \sim \frac{\pi}{k_F} \ \frac{\sin \left( 2k_FD + \varphi \right)}{D^2}\ 
{\rm e}^{-D/\lambda} ,
\end{equation}
with
\begin{equation}\label{eq:phase-shift}
\varphi \equiv \arctan \left( \frac{-1}{2k_F\lambda}\right) \approx 
\frac{-1}{2k_F\lambda}
\end{equation}
Thus, the presence of impurities in the non-magnetic spacer layer leads to an 
exponential decay of the interlayer exchange coupling with the distance between 
magnetic layers; this result is in contrast to the one obtained for the exchange 
interaction between magnetic ions in the previous section. It is traced back to 
the self-averaging character of the interaction between planes. Another effect 
of impurity scattering is the phase shift $\varphi$ 
[Eq.~(\ref{eq:phase-shift})].

The results of the present subsection are in full agreement with the ones 
obtained previously by Bruno {\em et al.\/} by using first-principles 
calculations together with the ``vertex cancellation theorem.''\cite{ Bruno1996, 
Bruno1997} Moreover, this previous study allows us to generalize the above 
result to the case of a spacer material with non-spherical Fermi surface; in 
this case, one obtains 
\begin{equation}
I(D) \sim \sum_\alpha \ I_\alpha \ \frac{\sin\left( q_{\bot \alpha}D + 
\phi_\alpha \right)}{D^2}\ {\rm e}^{-D/\lambda_\alpha} .
\end{equation}
In the above equation, $q_{\bot \alpha}$ and $\lambda_\alpha^{-1}$ are the real 
and imaginary parts of stationary spanning vectors of the {\em complex Fermi 
surface\/}\cite{ Bruno1995} of the alloy spacer material (since there may be 
several such vectors, they are labeled by the index $\alpha$);  $I_\alpha$ and 
$\phi_\alpha$ are the corresponding amplitude and phase. 

The model calculation presented here provides a simple physical explaination for 
the ``vertex cancellation theorem.'' As explained at the beginning of 
Sec.\ref{sec:ions}, only the paths going in straight line between two ions 
contribute significantly to the exchange interaction between them; this, 
together with the self-averaging property for the coupling between layers, forms 
the physical basis of the ``vertex cancellation theorem.''

\subsection{Effect of interface roughness}\label{sec:rough}

Next we discuss the effect of interface roughness on the interlayer exchange 
coupling. To be specific, we consider the case where the normal coordinates 
$r_{\bot 1}$ and $r_{\bot 2}$ characterizing $F_1$ and $F_2$ vary with the 
in-plane coordinates ${\bf r}_{\| 1}$ and ${\bf r}_{\| 2}$. The roughness is 
characterized by at least two parameters: the average amplitude of the 
fluctuations of $r_{\bot 1}$ and $r_{\bot 2}$, and the lateral correlation 
length, $\xi$.

The simplest approach to the effect of roughness consists in calculating the 
effective interlayer exchange coupling by averaging over thickness 
fluctuations.\cite{ Wang1990, Bruno1991, Bruno1992} In order to be allowed to 
do so however, some conditions must be satisfied. The first condition is that the 
lateral correlation length of the roughness, $\xi$, should be large enough for 
the interlayer exchange coupling to be locally well defined; typically this 
requires that $\xi > D$.\cite{ Bruno1992} On the other hand, we wish to consider 
that the sheets of spin are uniformly magnetized; but local fluctuations of 
spacer layer thickness induce local interlayer coupling fluctuations, that tend 
to produce local fluctuations of the magnetization direction in the magnetic 
layers. Thus, in order to keep the magnetization direction constant in the 
magnetic layers, the intralayer exchange coupling must be large enough, and the 
correlation length $\xi$ small enough; in practice, this condition $\xi$ is not 
very restrictive, and we shall not consider it further.

Within the above conditions, the effective interlayer exchange coupling 
$\overline{I}$ is given by
\begin{equation}\label{eq:average}
\overline{I} = \int\! dD \ P(D) \ I(D) ,
\end{equation}
where $P(D)$ is the distribution function of spacer thicknesses.\cite{ note} 
Thus, we have
\begin{equation}
\overline{I} \sim 2\pi\, \mbox{Re} \left[ \frac{-{\rm i}}{\gamma}\ \int\! dD\ 
P(D)\  \frac{{\rm e}^{{\rm i}\gamma D}}{D^2} \right] ;
\end{equation}
if the width of the distribution of thicknesses is small compared to the average 
thickness $D$, then the above equation becomes
\begin{equation}
\overline{I} \sim 2\pi\, \mbox{Re} \left[ \frac{-{\rm i}}{\gamma} \ A(\gamma )\  
\frac{{\rm e}^{{\rm i}\gamma \overline{D}}}{\overline{D}^2} \right]
\end{equation}
where
\begin{equation}\label{eq:A}
A(\gamma ) \equiv \int \! dD\ P(D)\ {\rm e}^{{\rm i}\gamma \left( 
D-\overline{D}\right)}
\end{equation}
is the form factor for the roughness. For a Gaussian distribution of width 
$\sigma$, i.e.,
\begin{equation}
P(D) = \frac{1}{\sqrt{2\pi}\sigma}\ {\rm e}^{-\left( D - \overline{D} 
\right)^2/2\sigma^2} ,
\end{equation}
one has
\begin{equation}
A(\gamma ) \approx {\rm e}^{-2{k_F}^2 \sigma^2} \ {\rm e}^{{\rm i}2k_F 
\sigma^2/\lambda } .
\end{equation}
Thus, the effective coupling becomes
\begin{equation}
\overline{I} \sim \frac{\pi}{k_F} \ {\rm e}^{-2{k_F}^2 \sigma^2} \ \frac{\sin 
\left( 2k_F \overline{D} + \varphi + \psi \right)}{\overline{D}^2}\ {\rm 
e}^{-\overline{D}/\lambda }
\end{equation}
where
\begin{equation}\label{eq:psi}
\psi \equiv \frac{2k_F \sigma^2}{\lambda} .
\end{equation}
The effect of interface roughness is essentially to attenuate the oscillatory 
coupling by the factor $\exp \left( -2k_F^2 \sigma^2 \right)$. In contrast to 
the attenuation due to impurities in the spacer layer, this effect is 
independent of the thickness of the spacer layer, $\overline{D}$. The phase 
shift $\psi$ in Eq.~(\ref{eq:psi}) is a {\em combined\/} effect of the roughness 
and of the impurities. An alternative approach for treating the effect of 
roughness is also presented in the Appendix.

In the more general case where the interlayer exchange coupling comprises 
several oscillatory components, the effect of the roughness is to strongly 
attenuate those oscillatory components which have a period of the order of (or 
smaller than) the amplitude of roughness, $\sigma$.\cite{ Wang1990, Bruno1991, 
Bruno1992}

\section{Conclusion}

By using a semi-classical approach, we have shown how the various kinds of 
defects, i.e., impurity scattering in the bulk of the layers and interface 
roughness, modify the exchange coupling between a pair of magnetic ions, or 
between two magnetic sheets. Our approach emphasizes the intrinsic 
self-averaging character of the coupling between layers, by contrast to the 
interaction between ions.  

We stress that although the same type of defects (impurities, roughness) play 
an essential r\^ole both in the interlayer exchange coupling and in the transport 
properties of magnetic multilayers, one cannot draw any simple relationship 
between their influence in the two cases. For example, in the absence of 
impurity scattering, it has been shown by Zhang and Levy\cite{ Zhang1991} that 
the effect of interface roughness on the perpendicular transport can be 
described in terms of an {\em effective mean free path\/}; but, as the above 
discussion has shown, the interlayer exchange coupling is not exponentially 
damped by the roughness alone, and it would be wrong to believe that this {\em 
effective mean free path\/} is of any significance for the interlayer exchange 
coupling. 

Here, we wish to comment on the limitations and extensions of the results 
obtained in this paper. For the sake of clarity and simplicity, we have 
restricted ourselves here to a very simple model (magnetic layers of 
infinitesimal thickness), and to a perturbative approach (expressing the 
exchange interaction via a susceptibility). In order to treat more realistic 
systems, with magnetic layers of finite (or infinite) thickness, a more 
sophisticated approach, such the one developed by Bruno,\cite{ Bruno1995} 
should be used; however, we expect that the conclusions obtained here would 
still hold. The effect of impurity scattering is described in terms of {\em 
complex wavevectors\/} and {\em complex Fermi surface\/}.\cite{ 
Bruno1996, Bruno1997} Impurity scattering in the spacer layer gives rise to an 
exponential damping of the coupling, but {\em not\/} impurity scattering in the 
magnetic layers (in this case, only the amplitudes and phases are affected), 
which agrees with what one would expect intuitively. Interface roughness 
modifies not only the spacer layer thickness, but also the thickness of the 
magnetic layers. It is known that the interlayer exchange coupling varies not 
only with the spacer thickness, but also with the thickness of the magnetic 
layer;\cite{ Bruno1995} however, the latter is a secondary effect and can be 
neglected here. Thus, we expect that the effect of roughness would be 
essentially the same as described within the simple approach of the present 
paper. 

Finally, we have considered only the limit case of ``geometrical'' roughness. An 
opposite limit case is the one of an interdiffusion of the magnetic material and 
spacer material near the interface. In such a situation, it is completey 
inappropriate to discuss the effect of roughness in terms of fluctuations of the 
spacer thickness. Rather, as done by Kudrnovsk\'y {\em et al.\/},\cite{ 
Kudrnovsky1996} one can consider that there is a thin layer of disordered 
magnetic alloy in the interface region, whose magnetization remains parallel to 
the magnetization of the magnetic layer nearby. In such a case, as one would 
expect intuitively, one finds that the interdiffusion modifies the amplitude(s) 
and phase(s) of oscillatory coupling, but does not alter the period(s) nor the 
$D^{-2}$ decay.

\section*{Acknowledgements}

P.M.L. acknowledges support by the Office of Naval Research, N00014-96-1-0203,
together with the Defense Advanced Research Projects Agency, N00014-96-1-1207, and 
the Japan Society for the Promotion of Science for his stay at the Institute for 
Materials Research in Sendai. He also thanks Albert Fert and the Laboratoire de 
Physique des Solides in Orsay, France, for their hospitality during his sabbatical 
stay there. P.B. gratefully acknowledges the hospitality of Junichiro Inoue at 
Nagoya University and the financial support of the Japan Society for the Promotion 
of Science.

\appendix
\section*{}

As we have shown in Sec.~\ref{sec:rough} interface roughness makes it necessary 
for us to average the interlayer coupling over a distribution of spacer thicknesses; 
see Eq.~\ref{eq:average}. Here, we present an alternate way of arriving at the 
attenuation factor due to 
roughness (thickness fluctuations) by using the canonical transformation 
introduced by Te\v{s}anovi\'c, Jari\'c and Maekawa to transform a film with a
rough boundary to one with smooth ones,\cite{ Tesanovic1986} and extended to films
with two rough boundaries by Meyerovich and Stepaniants.\cite{ Meyerovich1995} 
In doing this the 
transformation induces a perturbation (scattering potential) into an otherwise 
impurity free layer. It follows that we can replace the average over spacer layer 
thicknesses by one over a fluctuating phase shift that is induced by the canonical 
transformation that replaces the spacer with rough boundaries (interface roughness) 
by one with smooth boundaries. 

By following Refs.~{\onlinecite{ Tesanovic1986, Meyerovich1995} or  Trivedi 
and Ashcroft\cite{ Trivedi1988} the perturbation due to the surface roughness is
\begin{eqnarray}\label{eq:Vsurf}
V_{\rm surface}({\bf r}) &=& \frac{{\rm i}}{2\hbar} \eta ({\bf r}_\| ) \left\{ 
\left[ r_\bot p_\bot + p_\bot r_\bot \right] H_0 - {\rm c.c.} \right\} \nonumber \\
&&+2\eta ({\bf r}_\| ) H_0
\end{eqnarray}
where
\begin{eqnarray}
H_0 &\equiv& \frac{p_\bot^2}{2m} + V_0(\bot) + \frac{{\bf p}_\|^2}{2m} , \\
\eta ({\bf r}_\| ) &\equiv & \frac{\delta D({\bf r}_\| )}{\overline{D}} ,
\end{eqnarray}
and $V_0(r_\bot )$ is the confining potential, $\delta D({\bf r}_\| )$ is the 
variation in thickness over the surface, and $\overline{D}$ is the average 
thickness of the layer.

As our treatment up till now has neglected the confining potential, or 
equivalently in the limit of large $\overline{D}$, the distribution of 
eigenvalues is quasi-continuous. The matrix element of the perturbation, 
Eq.~(\ref{eq:Vsurf}), between states of $H_0$ at the Fermi level is
\begin{equation}
\left< {\bf k}_\| ,n \left| V_{\rm surface} \right| {\bf k}_\| + {\bf q}_\| ,n 
\right> = 2 \tilde\eta ({\bf q}_\| ) \varepsilon_F
\end{equation}
where $n$ labels the states referring to energy levels in the $r_\bot$ 
direction, and $\tilde\eta ({\bf q}_\| )$ is the two-dimensional Fourier 
transform of $\eta ({\bf r}_\| )$. 

By placing this scattering potential in the semiclassical expression for the 
Green's function\cite{ Bulaevskii1986}, we find the exchange interaction between 
two spins located on $F_1$ and $F_2$ is given again by Eq.~(\ref{eq:J}), but 
with a fluctuating  phase shift, Eq.~(\ref{eq:phi}), of
\begin{equation}\label{eq:phi-fluct}
\phi({\bf r}_1, {\bf r}_2) = 2k_F R_{1,2}\ \eta ({\bf r}_{\| 1} - {\bf r}_{\| 
2})  .
\end{equation}
By proceeding as in Secs.~\ref{sec:ions} and \ref{sec:sheets}, we obtain an 
expression of the interlayer coupling averaged over thickness fluctuations:
\begin{equation}
\overline{I} \sim 2\pi\, \mbox{Re} \left[ \frac{-{\rm i}}{2k_F} \ A^\prime 
(2k_F)\  \frac{{\rm e}^{{\rm i}2k_F \overline{D}}}{\overline{D}^2} \right] ,
\end{equation}
with 
\begin{equation}
A^\prime (2k_F ) \equiv \left< {\rm e}^{{\rm i}\phi ({\bf r}_\| )} \right> =  
\left< {\rm e}^{{\rm i}2k_F\, \delta D  ({\bf r}_\| ) } \right> .
\end{equation}
As in Eq.~(\ref{eq:shift}) we include a shift $\delta k$ by redefining $k_F$. 
Comparison with Eq.~(\ref{eq:A}) shows that $A^\prime (2k_F)$ is equal to 
$A(2k_F)$. Thus, one obtains results that are equivalent to those obtained in 
Sec.~\ref{sec:sheets}, when no impurities are present, i.e.,
$\lambda^{-1} = 0$.

\end{document}